\documentclass[11pt]{article}

\usepackage{mathptmx}
\usepackage{helvet}
\usepackage{courier}
\usepackage{pifont}
\usepackage{booktabs}
\usepackage{amsmath,amsfonts,amssymb,nicefrac}
\usepackage{algorithm,algorithmic}
\usepackage{tabularx,multirow}
\usepackage{graphicx}
\usepackage{subfigure,color,epsfig}

\newcommand\qedblob{\mbox{\ding{113}}}

\def\quantsymb{Q}

\newcommand{\cale}{{\cal E}}
\newcommand{\condition}{\,{\mbox{\large$|$}}\:}

\newcommand{\littlep}{{p}}

\newcommand{\sigmatwo}{{\Sigma_2^{\littlep}}}

\newcommand{\sigmatwok}{{\Sigma_{2k}^{\littlep}}}

\newcommand{\pitwo}{{\Pi_2^{\littlep}}}

\newtheorem{theorem}{Theorem}

\newtheorem{corollary}[theorem]{Corollary}

\newtheorem{lemma}[theorem]{Lemma}

\newtheorem{claim}[theorem]{Claim}
\newtheorem{proposition}[theorem]{Proposition}

\newcommand{\manyonereducesto}{\ensuremath{\leq_{m}^{p}}}

\newcommand{\co}{\ensuremath{\mathrm{co}}}
\newcommand{\p}{\ensuremath{\mathrm{P}}}
\newcommand{\np}{\ensuremath{\mathrm{NP}}}
\newcommand{\conp}{\ensuremath{\mathrm{coNP}}}
\newcommand{\ph}{\ensuremath{\mathrm{PH}}}
\newcommand{\pspace}{\ensuremath{\mathrm{PSPACE}}}
\newcommand{\fp}{\ensuremath{\mathrm{FP}}}

\newcommand{\todo}[2]{\textbf{ToDo{~(#1}):} \textit{#2}--}
\newcommand{\OMIT}[1]{} %

\newcommand{\EP}[3]{
\smallskip
\begin{center}
\begin{tabularx}{0.98\columnwidth}{ll}
\toprule
\multicolumn{2}{c}{#1} \\
\midrule
{\bf Given:}   & \parbox[t]{0.82\columnwidth}{#2\vspace*{1mm}}  \\
{\bf Question:}& \parbox[t]{0.82\columnwidth}{#3\vspace*{.5mm}} \\ 
\bottomrule
\end{tabularx}
\end{center}
\medskip
}

\newcommand\sigmalevel[1]{\ensuremath{{\Sigma^p_{#1}}}}

\newcommand\pilevel[1]{\ensuremath{{\Pi^p_{#1}}}}

\def\literalqed{{\ \nolinebreak\hfill\mbox{\qedblob\quad}}}

\def\qed{\literalqed}
\newenvironment{proofs}{\noindent{\sc Proof.}}{\literalqed\medskip}

\newcommand{\sproofof}[1]{\noindent{\bf Proof of {#1}. }}

\newcommand{\eproofof}[1]{\hfill \mbox{\qed$_{\mbox{\small {#1}}}$}\medskip}

\newcommand{\qbfk}[1]{\ensuremath{\mathrm{QBF}_{#1}}}
\newcommand{\qbf}{\ensuremath{\mathrm{QBF}}}

\newcommand{\width}{\ensuremath{\mathit{width}}}
\newcommand{\blocks}{\ensuremath{\mathit{blocks}}}
\newcommand{\score}{\ensuremath{\mathit{score}}}
\newcommand{\maxscore}{\ensuremath{\mathit{maxscore}}}

\newcommand{\msa}{\ensuremath{\mathit{maxsatasg}}}
\newcommand{\vetoes}{\ensuremath{\mathit{vetoes}}}

\newcommand{\partition}{\mathrm{Partition}}

\newcommand{\systemucm}[1]{{\mathrm{{#1}}\hbox{-}\mathrm{UCM}}}

\newcommand{\systemwcm}[1]{{\mathrm{{#1}}\hbox{-}\mathrm{WCM}}}

\newcommand{\systemducm}[1]{{\mathrm{{#1}}\hbox{-}\mathrm{DUCM}}}

\newcommand{\systemdwcm}[1]{{\mathrm{{#1}}\hbox{-}\mathrm{DWCM}}}

\newcommand{\systemucmnonempty}[1]{{\mathrm{{#1}}\hbox{-}\mathrm{UC_{\neq \emptyset}M}}}

\newcommand{\systemwcmnonempty}[1]{{\mathrm{{#1}}\hbox{-}\mathrm{WC_{\neq \emptyset}M}}}

\newcommand{\systemducmnonempty}[1]{{\mathrm{{#1}}\hbox{-}\mathrm{DUC_{\neq \emptyset}M}}}

\newcommand{\systemdwcmnonempty}[1]{{\mathrm{{#1}}\hbox{-}\mathrm{DWC_{\neq \emptyset}M}}}

\newcommand{\onlineucmk}[1]{{\mathrm{online}\hbox{-}\cale\hbox{-}\mathrm{UCM}[{#1}]}}
\newcommand{\onlineucm}{{\mathrm{online}\hbox{-}\cale\hbox{-}\mathrm{UCM}}}
\newcommand{\onlinesystemucmk}[2]{{\mathrm{online}\hbox{-}\mathrm{{#1}}\hbox{-}\mathrm{UCM}[{#2}]}}
\newcommand{\onlinesystemucm}[1]{{\mathrm{online}\hbox{-}\mathrm{{#1}}\hbox{-}\mathrm{UCM}}}
\newcommand{\onlinewcmk}[1]{{\mathrm{online}\hbox{-}\cale\hbox{-}\mathrm{WCM}[{#1}]}}
\newcommand{\onlinewcm}{{\mathrm{online}\hbox{-}\cale\hbox{-}\mathrm{WCM}}}
\newcommand{\onlinesystemwcmk}[2]{{\mathrm{online}\hbox{-}\mathrm{{#1}}\hbox{-}\mathrm{WCM}[{#2}]}}
\newcommand{\onlinesystemwcm}[1]{{\mathrm{online}\hbox{-}\mathrm{{#1}}\hbox{-}\mathrm{WCM}}}

\newcommand{\onlinesystemducmk}[2]{{\mathrm{online}\hbox{-}\mathrm{{#1}}\hbox{-}\mathrm{DUCM}[{#2}]}}
\newcommand{\onlinesystemducm}[1]{{\mathrm{online}\hbox{-}\mathrm{{#1}}\hbox{-}\mathrm{DUCM}}}

\newcommand{\onlinesystemdwcmk}[2]{{\mathrm{online}\hbox{-}\mathrm{{#1}}\hbox{-}\mathrm{DWCM}[{#2}]}}
\newcommand{\onlinesystemdwcm}[1]{{\mathrm{online}\hbox{-}\mathrm{{#1}}\hbox{-}\mathrm{DWCM}}}

\newcommand{\pinpointonlinesystemucmk}[2]{{\mathrm{pinpoint}\hbox{-}\mathrm{online}\hbox{-}\mathrm{{#1}}\hbox{-}\mathrm{UCM}[{#2}]}}

\newcommand{\schedulerobustonlinesystemucmk}[2]{{\mathrm{SR}\hbox{-}\mathrm{online}\hbox{-}\mathrm{{#1}}\hbox{-}\mathrm{UCM}[{#2}]}}
\newcommand{\schedulerobustonlinesystemucm}[1]{{\mathrm{SR}\hbox{-}\mathrm{online}\hbox{-}\mathrm{{#1}}\hbox{-}\mathrm{UCM}}}

\newcommand{\ehnote}[1]{}
\newcommand{\lahnote}[1]{}
\newcommand{\jrnote}[1]{}

 \title{The Complexity of Online Manipulation of Sequential
Elections\thanks{Also appears as URCS-TR-2012-974.}
}

\author{Edith Hemaspaandra\thanks{Supported in part by grant
   NSF-CCF-1101452
   and a
   Friedrich Wilhelm Bessel Research Award.
   Work done
   in part while visiting Heinrich-Heine-Universit\"{a}t D\"{u}sseldorf.} \\
        Department of Computer Science \\
        Rochester Institute of Technology \\
        Rochester, NY 14623, USA 
\and
        Lane A. Hemaspaandra\thanks{Supported in part by grants
   NSF-CCF-\{0915792,\allowbreak{}1101479\}
   and \mbox{ARC}-\mbox{DP}110101792, and a
   Friedrich Wilhelm Bessel Research Award.
   Work done
   in part while visiting Heinrich-Heine-Universit\"{a}t D\"{u}sseldorf.} \\ 
        Department of Computer Science \\
        University of Rochester \\
        Rochester, NY 14627, USA
\and
        J{\"o}rg Rothe\thanks{Supported in part by DFG grant RO-1202/15-1,
 SFF grant ``Cooperative Normsetting'' of HHU D{\"u}sseldorf,
 \mbox{ARC}-\mbox{DP}110101792, and a DAAD grant for a PPP project in the PROCOPE
 program.} \\
        Institut f\"ur Informatik \\
        Heinrich-Heine-Universit{\"a}t D{\"u}sseldorf  \\
        40225 D\"usseldorf, Germany
}
\date{February 29, 2012; revised May 12, July 5, and Sept.\ 28, 2012}

\begin{document}
\sloppy

\maketitle
\begin{abstract}  

  Most work on manipulation assumes that all preferences are known to
  the manipulators.  However, in many settings elections are open and
  sequential, and manipulators may know the already cast votes but may
  not know the future votes.  We introduce a framework, in which
  manipulators can see the past votes but not the future ones, to
  model online coalitional manipulation of sequential elections, and we
  show that in this setting manipulation can be extremely complex even
  for election systems with simple winner problems.  Yet we also
  show that for some of the most important election systems such
  manipulation is simple in certain settings.  This suggests that when
  using sequential voting, one should pay great attention to the
  details of the setting in choosing one's voting rule.

  Among the highlights of our classifications are: We show that,
  depending on the size of the manipulative coalition, the 
  online manipulation problem can be complete for 
  each level of the polynomial
  hierarchy or even for $\pspace$. 
  We obtain the most dramatic contrast to date between
  the nonunique-winner and unique-winner models:
  Online weighted manipulation for plurality is
  in $\p$ in the nonunique-winner model, 
  yet is $\conp$-hard (constructive case)
  and $\np$-hard (destructive case) in the unique-winner model.
  And we obtain what to the best of our knowledge are the first 
$\p^{\np[1]}$-completeness and 
$\p^{\np}$-completeness results in the field of 
computational social choice, in particular proving 
such completeness for, respectively, the 
complexity 
of 3-candidate and 4-candidate (and unlimited-candidate) 
online weighted coalition manipulation of 
veto elections.
\end{abstract}

\section{Introduction}

Voting is a widely used method for preference aggregation and
decision-making.  In particular, \emph{strategic} voting (or
\emph{manipulation}) has been studied intensely in social choice
theory (starting with the celebrated work of
Gibbard~\cite{gib:j:polsci:manipulation} and
Satterthwaite~\cite{sat:j:polsci:manipulation}) and, in the rapidly emerging
area of \emph{computational} social choice, also with respect to its
algorithmic properties and computational complexity (starting with the
seminal work of
Bartholdi, Tovey, and Trick~\cite{bar-tov-tri:j:manipulating};
see the surveys~\cite{fal-hem-hem:j:cacm-survey,fal-hem-hem-rot:b:richer}).
This computational aspect is
particularly important in light of the many applications of voting in
computer science, ranging from meta-search heuristics for the internet
\cite{dwo-kum-nao-siv:c:rank-aggregation}, to recommender systems
\cite{gho-her-mun-sen:c:voting-for-movies} and multiagent systems in
artificial intelligence (see the survey by
Conitzer~\cite{con:j:making-decisions}).

Most of the previous work on manipulation, however, is concerned with
voting where the manipulators know the nonmanipulative
votes.
Far less attention has been paid 
(see the related work
below) to manipulation 
in the midst of elections that are 
modeled as dynamic processes.

We introduce a novel framework for online manipulation,
where voters vote in sequence and the current manipulator, who knows
the previous votes and which voters are still to come but does not
know their votes, must decide---%
right at that moment---what the ``best'' vote to cast
is.  So, while other approaches to sequential
voting are game-theoretic,
stochastic,
or axiomatic in nature (again, see the related work), our approach
to manipulation of sequential voting
is shaped by the area of 
``online algorithms''
\cite{bor-ely:b:online-algorithms}, in the technical sense of a setting
in which one (for us, 
each manipulative voter) 
is being asked to make a manipulation decision just on the basis of
the information one has in one's hands at the moment even though
additional information/system evolution may well be happening down the
line.  In this area, there are different frameworks for evaluation.
But the most attractive one, which pervades the area as a general
theme, is the idea that one may want to ``maxi-min''
things---\emph{one may want to take the action that
  maximizes the goodness of the set of outcomes that one can expect
  regardless of what happens down the line from one time-wise}.  For
example, if the current manipulator's preferences are $\rm Alice > Ted
> Carol > Bob$ and if she can cast a (perhaps insincere) vote that
ensures that Alice or Ted will be a winner no matter what later voters
do, and there is no vote she can cast that ensures that Alice will
always be a winner, this maxi-min approach would say that that vote is
a ``best'' vote to cast.

It will perhaps be a bit surprising to those familiar with online
algorithms and competitive analysis that in our model of online
manipulation we will not use a (competitive) \emph{ratio}.  The reason
is that voting commonly uses an \emph{ordinal} preference model, in
which preferences are total orders of the candidates.  It would be a
severely improper step to jump from that to assumptions about
intensity of preferences and utility, e.g., to assuming that everyone
likes her $n$th-to-least favorite candidate exactly $n$ times
more than she likes her least favorite candidate.

\paragraph{Related Work.}
Conitzer and Xia~\cite{con-xia:c:stackelberg-sequential}
(see also the related paper by
Desmedt and Elkind \cite{des-elk:c:sequential-voting})
define and study the
Stackelberg voting game (also quite naturally called, in an earlier
paper that mostly looked at two candidates, the roll-call voting
game~\cite{slo:j:sequential-voting}).  This basically is an election
in which the voters vote in order, \emph{and the preferences are
  common knowledge---everyone knows everyone else's preferences,
  everyone knows that everyone knows everyone else's preferences, and
  so on out to infinity}.  Their analysis of this game is
fundamentally game-theoretic;
with such complete
knowledge in a sequential setting, there is precisely one (subgame
perfect Nash) equilibrium, which can be computed from the back end
forward.  Under their work's setting and assumptions, for bounded 
numbers of manipulators manipulation is in P, but we will show that
in our model even with bounded numbers of manipulators 
manipulation sometimes (unless 
$\p = \np$) falls beyond $\p$.

The interesting ``dynamic voting'' work of
Tennenholtz~\cite{ten:c:transitive-voting}
investigates sequential voting,
but focuses on axioms and voting rules rather than
on coalitions and manipulation.
Much heavily Markovian work
studies sequential decision-making and/or
dynamically varying preferences;
our work in contrast is
nonprobabilistic and focused on
the complexity of coalitional manipulation.
Also somewhat related to, but quite 
different from, our work is the work on possible
and necessary winners.  The seminal paper on that is due to
Konczak and Lang~\cite{kon-lan:c:incomplete-prefs},
and more recent work includes
\cite{con-xia:c:possible-necessary-winners,bet-hem-nie:c:parameterized-possible-winner,bac-bet-fal:c:probabilistic-possible-winner,bet:c:problem-kernels-possible-winner,bet-dor:j:possible-winner-dichotomy,che-lan-mau-mon-xia:j:possible-winners-new-candidates-scoring,bau-rot:j:possible-winner-dichotomy-final-step,lan-pin-ros-sal-ven-wal:j:winner-in-voting-trees-incomplete-prefs%
};  
the biggest difference is that those are, loosely, one-quantifier
settings, but the more dynamic setting of online manipulation 
involves numbers of quantifiers that can grow with the input size.
Another related research line studies multi-issue elections
\cite{con-xia:c:strategyproof-multiissue,con-lan-xia:c:mle-multiissue,con-lan-xia:c:sequential-multiissue-paradoxes,con-lan-xia:c:hypercubewise-multiissue-elections};
although there the separate issues may run in
sequence, each issue typically is voted on simultaneously and
with preferences being common knowledge.

\paragraph{Organization.}

We first provide the needed preliminaries for 
(standard and sequential) 
elections, manipulation, and scoring rules, and give some background
from complexity theory.  Then, after introducing our model of
online manipulation 
formally, we will present some general
complexity results on the problems defined, and also some specific
results for online manipulation
in natural voting systems (i.e., for some
central scoring rules).  Finally, we turn to schedule-robust online
manipulation, a setting in which not even the order of future voters
is known to the current manipulator.

\section{Preliminaries}

\paragraph{Elections.}
A \emph{(standard, i.e., simultaneous) election} 
$(C,V)$ is specified by a set $C$ of
candidates and a list~$V$, where we assume that each element in $V$ is
a pair $(v,p)$ such that $v$ is a voter name and $p$ is $v$'s vote.
How the votes in $V$ are represented depends on the election system
used---we assume, as is required by most systems, votes to be total
preference orders over~$C$.
For example, if $C = \{a,b,c\}$, a vote
of the form $c>a>b$ means that this voter (strictly) prefers $c$ to
$a$ and $a$ to~$b$.

We introduce election snapshots to capture 
sequential election
scenarios as follows.
Let $C$ be a set of candidates and
let 
$u$
be 
(the name of) a voter.
An \emph{election snapshot for $C$ and $u$} is specified by a triple $V =
(V_{<u}, u, V_{u<})$ consisting of all voters
in the order they
vote, along with, for each voter before~$u$
(i.e., those in $V_{<u}$), the vote she cast,
and for each voter after~$u$ (i.e., those
in $V_{u<}$), a bit specifying if she is part of the manipulative
coalition (to which $u$
always belongs).
That is, $V_{<u} = ((v_1,p_1), (v_2,p_2), \ldots,
(v_{i-1},p_{i-1}))$, where the voters named $v_1, v_2, \ldots,
v_{i-1}$ (including perhaps manipulators and nonmanipulators) have
already cast their votes (preference order $p_j$ being cast by~$v_j$),
and $V_{u<} = ((v_{i+1},x_{i+1}),
(v_{i+2},x_{i+2}), \ldots, (v_n,x_n))$ lists the names of the voters
still to cast their votes, in that order, and where $x_j = 1$ if $v_j$
belongs to the manipulative coalition and $x_j = 0$ otherwise.

\vspace*{-3mm}
\paragraph{Scoring Rules.}
A \emph{scoring rule} for $m$ candidates is given by a scoring vector
$\alpha = (\alpha_1, \alpha_2, \ldots, \alpha_m)$ of nonnegative
integers such that $\alpha_1 \geq \alpha_2 \geq \cdots \geq \alpha_m$.
For an election $(C,V)$, each candidate $c \in C$ scores $\alpha_i$
points for each vote that ranks $c$ in the $i$th position.  Let
$\score(c)$ be the total score of $c \in C$.  All candidates scoring
the most points are winners of $(C,V)$.  Some of the most popular
voting systems are \emph{$k$-approval} (especially \emph{plurality}, 
aka 1-approval) 
and \emph{$k$-veto} (especially \emph{veto}, aka 1-veto).
Their $m$-candidate, $m\geq k$, versions are
defined by the scoring 
vectors
$(\underbrace{1, \ldots , 1}_{k}, \underbrace{0, \ldots , 0}_{m-k})$ and
$(\underbrace{1, \ldots , 1}_{m-k}, \underbrace{0, \ldots , 0}_{k})$.
When $m$ is not fixed, we omit the phrase ``$m$-candidate.''

\paragraph{Manipulation.}
The
\emph{(standard) weighted coalitional manipulation 
problem}~\cite{con-lan-san:j:when-hard-to-manipulate},
abbreviated by $\systemwcm{\cale}$, for any election system $\cale$ is
defined as follows:
\EP{$\cale$-Weighted-Coalitional-Manipulation}
{A candidate set $C$, a list $S$ of nonmanipulative voters each
having a
nonnegative
integer weight, a list $T$ of the nonnegative integer 
weights of the
manipulative voters (whose preferences over $C$ are unspecified),
with $S \cap T = \emptyset$, and a distinguished candidate $c \in C$.}
{Can the manipulative votes $T$ be set such that $c$ is a (or the) $\cale$
winner of $(C,S \cup T)$?}

Asking whether $c$ can be 
made ``a winner'' is called the nonunique-winner model and is the 
model of all notions in this paper unless mentioned otherwise.
If one asks whether $c$ can be made a ``one and only winner,''
that is called the unique-winner model.
We also use the \emph{unweighted} variant,
where each vote has unit weight, and write $\systemucm{\cale}$ as a
shorthand.  Note that $\systemucm{\cale}$ with a \emph{single}
manipulator (i.e., $\|T\| = 1$ in the problem instance) is the
manipulation problem originally studied in
\cite{bar-tov-tri:j:manipulating,bar-orl:j:polsci:strategic-voting}.
Conitzer, Sandholm, and
Lang~\cite{con-lan-san:j:when-hard-to-manipulate} also introduced the
\emph{destructive} variants of these manipulation problems, where the
goal is not to make $c$ win but to 
ensure that $c$ is not a winner, and we
denote the corresponding problems by $\systemdwcm{\cale}$ and
$\systemducm{\cale}$.  Finally, we write $\systemwcmnonempty{\cale}$,
$\systemucmnonempty{\cale}$, $\systemdwcmnonempty{\cale}$, and
$\systemducmnonempty{\cale}$ to indicate that the problem instances
are required to have a nonempty coalition of manipulators.

\paragraph{Complexity-Theoretic Background.}  
We assume the reader is familiar with basic complexity-theoretic
notions such as the complexity classes $\p$ and $\np$, the class $\fp$
of polynomial-time computable functions, polynomial-time many-one
reducibility ($\manyonereducesto$), and hardness and completeness with
respect to $\manyonereducesto$ for a complexity class (see, e.g., the
textbook~\cite{pap:b:complexity}).

Meyer and Stockmeyer~\cite{mey-sto:c:reg-exp-needs-exp-space} and
Stockmeyer~\cite{sto:j:poly}
introduced and studied the polynomial hierarchy, $\ph = \bigcup_{k\geq
  0} \sigmalevel{k}$, whose levels are inductively defined by
$\sigmalevel{0} = \p$ and $\sigmalevel{k+1} = \np^{\sigmalevel{k}}$,
and their co-classes, $\pilevel{k} = \co\sigmalevel{k}$ for $k \geq 0$.
They also characterized these levels by polynomially length-bounded
alternating existential and universal quantifiers.

\begin{lemma}[Meyer and Stockmeyer~\cite{mey-sto:c:reg-exp-needs-exp-space}
 and Stockmeyer~\cite{sto:j:poly}]
\label{lem:alternating-quantifiers}
For all $k \geq 0$, $X \in \sigmalevel{k}$ if and only if there exist
a set $Y \in \p$ and a polynomial $p$ such that for each input~$x$,
$x \in X$ if and only if
\[
(\exists^p z_1)\, (\forall^p z_2)\, \cdots\, (\quantsymb^p z_k)\,
[(x,z_1, z_2, \ldots , z_k) \in Y],
\]
where $(\exists^p z_i)$ stands for $(\exists z_i: |z_i| \leq p(|x|))$,
$(\forall^p z_i)$ for $(\forall z_i: |z_i| \leq p(|x|))$,
$\quantsymb^p = \exists^p$ if $k$ is odd, and $\quantsymb^p =
\forall^p$ if $k$ is even.

For each $k \geq 0$, $\pilevel{k}$ is characterized
analogously by switching $\exists^p$ and~$\forall^p$.
\end{lemma}

$\p^{\np}$ is the class of problems solvable in deterministic
polynomial time with access to an $\np$ oracle, and $\p^{\np[1]}$ is
the restriction of $\p^{\np}$ where only one oracle query is allowed.
Note that
\[
\p \subseteq \np \cap \conp \subseteq \np \cup \conp \subseteq
\p^{\np[1]} \subseteq \p^{\np} \subseteq \sigmalevel{2} \cap
\pilevel{2} \subseteq \sigmalevel{2} \cup \pilevel{2} \subseteq \ph
\subseteq \pspace,
\]
where $\pspace$ is the class of
problems solvable in polynomial space.  
The \emph{quantified boolean formula
problem}, $\qbf$, is a standard $\pspace$-complete
problem.
$\qbfk{k}$ ($\widetilde{\mathrm{QBF}}_{k}$)
denotes the restriction of $\qbf$
with at most $k$ 
quantifiers that start
with $\exists$ ($\forall$) and then alternate between $\exists$
and~$\forall$, and we assume that each $\exists$ and $\forall$
quantifies over a set of boolean variables.  For each $k \geq 1$,
$\qbfk{k}$ is $\sigmalevel{k}$-complete and $\widetilde{\mathrm{QBF}}_{k}$ is
$\pilevel{k}$-complete~\cite{mey-sto:c:word-exponential,wra:j:complete}.

\section{Our Model of Online Manipulation}

The core of our model of online manipulation in 
sequential voting is what we call
the \emph{magnifying-glass moment}, namely, the moment at which a
manipulator $u$ is the one who is going to vote, is aware of what has
happened so far in the election (and which voters are still to come,
but in general not knowing what they want, except in the case of
voters, if any, who are coalitionally linked to~$u$).
In this moment, $u$ seeks to ``figure out'' what the ``best'' vote to
cast is.  We will call
the information available in such a moment an
\emph{online manipulation setting} (\emph{OMS}, for short) and define
it formally as a tuple $(C, u, V, \sigma, d)$, where $C$ is a set
of candidates; $u$ is a distinguished voter; $V = (V_{<u}, u,
V_{u<})$ is an election snapshot for $C$ and~$u$; $\sigma$ is the
preference order of the manipulative coalition to which $u$ belongs;
and $d \in C$ is a distinguished candidate.  
Given an election system~$\cale$, define the \emph{online unweighted
  coalitional manipulation problem}, abbreviated by
$\onlinesystemucm{\cale}$, as follows:
\EP{online-$\cale$-Unweighted-Coalitional-Manipulation}
{An OMS $(C, u, V, \sigma, d)$ as described above.}
{Does there exist some vote that $u$ can cast (assuming support from
  the
  manipulators coming after $u$) such that no matter what votes are
  cast by the nonmanipulators coming after $u$, there exists some
  $c \in C$ such that $c \geq_{\sigma} d$ and $c$ is an $\cale$ winner of
  the election?}

By ``support from the manipulators coming after $u$'' we mean that $u$'s
coalition partners coming after $u$, when they get to vote, will use
their then-in-hand knowledge of all votes up to then to help $u$ reach
her goal: By a joint effort $u$'s coalition can ensure that the
$\cale$ winner set will always include a candidate liked by the
coalition as much as or more than~$d$, even when the nonmanipulators
take their strongest action so as to prevent this.  Note that this
candidate, $c$ in the problem description, may be different based on
the nonmanipulators' actions.  (Nonsequential manipulation problems 
usually focus on whether a single candidate can be made to win, 
but in our setting, this ``that person or better'' focus is more 
natural.)

For the case of weighted manipulation, each voter also comes with a 
nonnegative integer weight.
We denote this problem by $\onlinesystemwcm{\cale}$.

We write $\onlinesystemucmk{\cale}{k}$ in the
unweighted case and $\onlinesystemwcmk{\cale}{k}$ in the weighted case
to denote the
problem when the number of manipulators from $u$
onward is restricted to be at most $k$.

Our corresponding destructive problems are denoted by
$\onlinesystemducm{\cale}$, $\onlinesystemdwcm{\cale}$,
$\onlinesystemducmk{\cale}{k}$, and $\onlinesystemdwcmk{\cale}{k}$.
In $\onlinesystemducm{\cale}$ we ask whether the given
current manipulator $u$ (assuming support from the manipulators after
her) can cast a vote such that no matter what votes are cast by
the nonmanipulators after $u$, no $c \in C$ with $d \geq_{\sigma} c$
is an $\cale$ winner of the election, i.e., $u$'s coalition can ensure
that the $\cale$ winner set never includes 
$d$ or any even more hated candidate.  The other three 
problems are defined
analogously.

Note that $\onlinesystemucm{\cale}$ generalizes the original
unweighted manipulation problem with a single manipulator as
introduced by
Bartholdi, Tovey, and Trick~\cite{bar-tov-tri:j:manipulating}.
Indeed, their manipulation problem in effect is the special case
of $\onlinesystemucm{\cale}$ when restricted to instances where 
there is just one manipulator, she is the last voter to cast a vote,
and $d$ is the coalition's most preferred candidate.
Similarly, $\onlinesystemwcm{\cale}$ generalizes the 
(standard) coalitional
weighted manipulation problem (for nonempty coalitions of
manipulators).
Indeed, that traditional manipulation problem is the special
case of $\onlinesystemwcm{\cale}$, restricted to instances 
where only manipulators come after $u$ and 
$d$ is the coalition's most preferred candidate.
If
we take an analogous approach except with
$d$ restricted now to being the most 
hated candidate of the coalition, 
we generalize the corresponding notions for the
destructive cases.  We summarize these observations as follows.

\begin{proposition}
  For each election system~$\cale$, it holds that
\begin{enumerate}
\item $\systemucmnonempty{\cale} \manyonereducesto \onlinesystemucm{\cale}$,
\item $\systemwcmnonempty{\cale} \manyonereducesto \onlinesystemwcm{\cale}$,
\item $\systemducmnonempty{\cale} \manyonereducesto \onlinesystemducm{\cale}$,
  and
\item $\systemdwcmnonempty{\cale} \manyonereducesto \onlinesystemdwcm{\cale}$.
\end{enumerate}
\end{proposition}

Corollary~\ref{cor:offline-to-online} below follows immediately from
the above proposition.

\begin{corollary}
\label{cor:offline-to-online}
\begin{enumerate}
\item
For each election system $\cale$ such that the (unweighted) winner
problem is solvable in polynomial time, it holds that
$\systemucm{\cale} \manyonereducesto \onlinesystemucm{\cale}$.
\item
For each election system $\cale$ such that the weighted winner
problem is solvable in polynomial time, it holds that
$\systemwcm{\cale} \manyonereducesto \onlinesystemwcm{\cale}$.
\item
For each election system $\cale$ such that the winner
problem is solvable in polynomial time, it holds that
$\systemducm{\cale} \manyonereducesto \onlinesystemducm{\cale}$.
\item
For each election system $\cale$ such that the weighted winner
problem is solvable in polynomial time, it holds that
$\systemdwcm{\cale} \manyonereducesto \onlinesystemdwcm{\cale}$.
\end{enumerate}
\end{corollary}

We said above that, by default, we will use the \emph{nonunique-winner model}
and all the above problems are defined in this model.  However, we
will also have some results in the \emph{unique-winner model}, which
will, here, sharply contrast with the corresponding results in the 
nonunique-winner
model.  To indicate that a problem, such as $\onlinesystemucm{\cale}$,
is in the unique-winner model, we write
$\onlinesystemucm{\cale}_{\mathrm{UW}}$ and ask whether the current
manipulator $u$ (assuming support from the manipulators coming 
after her)
can ensure that there exists some $c \in C$ such that $c \geq_{\sigma}
d$ and $c$ is \emph{the unique $\cale$ winner} of the election.

\section{General Results}

\begin{theorem}
\label{thm:general-pspace}  
\begin{enumerate}
\item For each election system $\cale$ whose weighted winner problem can be
  solved in polynomial time,\footnote{We mention in passing
  here, and henceforward we will not explicitly mention it
  in the analogous cases, that the claim clearly remains true
  even when
  ``polynomial time'' is replaced by the larger class
  ``polynomial space.''}
  the problem $\onlinewcm$ is in $\pspace$.
\item For each election system $\cale$ whose winner problem can be
  solved in polynomial time,
  the problem $\onlineucm$ is in $\pspace$.

\item There exists an election system $\cale$ with a polynomial-time
 winner problem such that the problem $\onlineucm$ is $\pspace$-complete.

\item There exists an election system $\cale$ with a polynomial-time
 weighted winner problem such that the problem $\onlinewcm$ is
$\pspace$-complete.
\end{enumerate}
\end{theorem}

\begin{proofs}
  The proof of the first statement (which is analogous to the proof of
  the first statement in Theorem~\ref{t:pptc}) follows from the easy
  fact that $\onlinewcm$ can be solved by an alternating Turing
  machine in polynomial time, and thus, due to the characterization of
  Chandra, Kozen, and Stockmeyer~\cite{cha-koz-sto:j:alternation}, by
  a deterministic Turing machine in polynomial space.
  The proof of the second case is analogous.

  We construct an
  election system $\cale$ establishing the third statement.
  Let $(C,u,V,\sigma,d)$ be a given input.
  $\cale$ will look at the lexicographically least
  candidate name in~$C$.  Let $c$ represent that name string in some
  fixed, natural encoding.  $\cale$ will check if $c$ represents a
  \emph{tiered} boolean formula, by which we mean one whose
  variable names are all of the form $x_{i,j}$ (which really means a
  direct encoding of a string, such as ``$x_{4,9}$''); the $i,j$
  fields must all be positive integers.  If $c$ does not represent
  such a tiered formula, everyone loses on that input.  Otherwise
  (i.e., if $c$ represents a tiered formula),
let $\width$ be the maximum $j$
  occurring as the second subscript in any variable name ($x_{i,j}$)
  in~$c$, and let $\blocks$ be the maximum $i$ occurring as the first
  subscript in any variable name in~$c$.  If there are fewer than
  $\blocks$ voters in~$V$, everyone loses.  Otherwise, if there are
  fewer than $1 + 2 \cdot \width $ candidates in~$C$, everyone
  loses (this is so that each vote will involve enough candidates that
  it can be used to set all the variables in one block).
  Otherwise, if there exists some~$i$, $1 \leq i \leq \blocks$, such
  that for no $j$ does the variable $x_{i,j}$ occur in~$c$, then
  everyone loses.  Otherwise, order the voters from the
  lexicographically least to the lexicographically greatest voter
  name. 
If distinct voters are allowed to have the same name string
(e.g., John Smith),
we break ties by sorting according to the
associated preference orders within each group of tied voters
(second-order ties are no problem, as those votes are identical, so
any order will have the same effect).
 Now, the first voter
  in this order will assign truth values to all
  variables~$x_{1,\ast}$, the second voter in this order will assign
  truth values to all variables~$x_{2,\ast}$, and so on up to the
  $\blocks$th voter, who will assign truth values to all
  variables~$x_{\blocks,\ast}$.

  How do we get those assignments from these votes?  Consider a vote
  whose total order over $C$ is $\sigma'$ (and recall
  that $\|C\| \geq 1 + 2 \cdot \width$).  Remove $c$ from
  $\sigma'$, yielding~$\sigma''$.  Let $c_1 <_{\sigma''} c_2 <_{\sigma''} \cdots  <_{\sigma''} c_{2 \cdot
    \width}$ be the $2 \cdot \width$ least preferred candidates in
  $\sigma''$.  We build a vector in $\{0,1\}^{\width}$ as follows:
  The $\ell$th bit of the vector is $0$ if the string that names
  $c_{1+2(\ell - 1)}$ is lexicographically less than the string that
  names $c_{2\ell}$, and this bit is $1$ otherwise.

  Let $b_i$ denote the vector thus built from the $i$th vote (in the above
  ordering), $1 \leq i \leq \blocks$.  Now, for each variable
  $x_{i,j}$ occurring in~$c$, assign to it the value of the $j$th bit
  of~$b_i$, where $0$ represents \emph{false} and $1$ represents
  \emph{true}.  We have now assigned all variables of~$c$, so $c$
  evaluates to either \emph{true} or \emph{false}.  If $c$ evaluates
  to \emph{true}, everyone wins, otherwise everyone loses.  This completes
  the specification of the election system~$\cale$.
  $\cale$ has a polynomial-time
winner problem, as any boolean formula,
  given an assignment to all its variables,
  can easily be evaluated in polynomial time.

  To show $\pspace$-hardness, we $\manyonereducesto$-reduce the
  $\pspace$-complete problem $\qbf$ to the problem $\onlineucm$.
  Let $y$ be an instance of $\qbf$.
  We transform $y$ into an instance of the form
\begin{align*}
& (\exists\, x_{1,1},
x_{1,2}, 
\ldots , x_{1,k_1})\,
(\forall\, x_{2,1},
x_{2,2}, 
\ldots , x_{2,k_2})\, \cdots
(\quantsymb_{\ell}\, x_{\ell,1},
x_{\ell,2}, 
\ldots , x_{\ell,k_{\ell}})\,
\\
& [\Phi(x_{1,1},
x_{1,2},
\ldots , x_{1,k_1},
x_{2,1},
x_{2,2},
\ldots , x_{2,k_2}, \ldots ,
x_{\ell,1},
x_{\ell,2}, 
\ldots , x_{\ell,k_{\ell}})]
\end{align*}
in polynomial time, where $\quantsymb_\ell = \exists$ if $\ell$ is odd
and $\quantsymb_\ell = \forall$ if $\ell$ is even, 
the $x_{i,j}$ are boolean variables,
$\Phi$ is a boolean formula,
and for each~$i$, $1 \leq i \leq \ell$, $\Phi$ contains at least one
variable of the form~$x_{i,\ast}$.  This quantified boolean formula is
$\manyonereducesto$-reduced to an instance $(C, u, V, \sigma, c)$ of
$\onlineucm$ as follows:
\begin{enumerate}
\item $C$ contains a candidate whose name, $c$, encodes~$\Phi$, and in
  addition $C$ contains $2 \cdot \max(k_1, \ldots , k_{\ell})$ other
  candidates, all with names lexicographically greater than~$c$---for
  specificity, let us say their names are the $2 \cdot \max(k_1,
  \ldots , k_{\ell})$ strings that immediately follow $c$ in
  lexicographic order.

\item $V$ contains $\ell$ voters, $1, 2, \ldots , \ell$, who vote in
  that order, where $u = 1$ is the distinguished voter and all odd
  voters belong to $u$'s manipulative coalition and all even voters do
  not.  The voter names will be lexicographically ordered by their
  number, $1$ is least and $\ell$ is greatest.

\item The manipulators' preference order $\sigma$ is to like
  candidates in the opposite of their lexicographic order.  In
  particular, $c$ is the coalition's most preferred candidate.
\end{enumerate}
This
is a polynomial-time reduction.  It follows immediately from this
construction and the definition of $\cale$ that $y$ is in $\qbf$ if
and only if $(C, u, V, \sigma, c)$ is in $\onlineucm$.

To prove the last statement, simply let $\cale$ be the election system that
ignores the weights of the voters and then works exactly as
the previous election system.~\end{proofs}

The following theorem
shows that for bounded numbers of manipulators the complexity 
crawls up the polynomial hierarchy.
The theorem's proof is based on the proof given above, 
except we need to use the alternating quantifier
characterization due to Meyer and
Stockmeyer~\cite{mey-sto:c:reg-exp-needs-exp-space} and
Stockmeyer~\cite{sto:j:poly} for the upper bound and to reduce from
the $\sigmatwok$-complete problem $\qbfk{2k}$ rather than from $\qbf$
for the lower bound.

\begin{theorem}
\label{t:pptc}  
Fix any $k \geq 1$.
\begin{enumerate}
\item For each election system $\cale$ whose weighted winner problem can be
  solved in polynomial time, the problem
  $\onlinewcmk{k}$ is in~$\sigmatwok$.
\item For each election system $\cale$ whose winner problem can be
  solved in polynomial time, the problem
  $\onlineucmk{k}$ is in~$\sigmatwok$.
\item
There exists an election system $\cale$
  with a polynomial-time winner problem such that the problem
  $\onlineucmk{k}$ is $\sigmatwok$-complete.
\item
There exists an election system $\cale$
  with a polynomial-time weighted winner problem such that the problem
  $\onlinewcmk{k}$ is $\sigmatwok$-complete.
\end{enumerate}
\end{theorem}

\begin{proofs}
  For the first statement, let $(C, u, V, \sigma, d)$ be an instance
  of $\onlinewcmk{k}$.  Let $u_1 = u$, and rename the manipulators
  after $u$ as $u_2, \ldots , u_k$ in the order they vote.
  Thus, $(C, u, V, \sigma, d)$ is in
  $\onlinewcmk{k}$ if and only if there exists some preference order
  for $u_1$ such that for all preference orders the nonmanipulators
  between $u_1$ and $u_2$ (if any) can cast, there exists some
  preference order for $u_2$ such that \ldots\ there exists some
  preference order for $u_k$ such that for all preference orders the
  nonmanipulators after $u_k$ (if any) can cast, the $\cale$ winner
  set under the votes just cast contains at least one candidate $c \in
  C$ such that $c \geq_{\sigma} d$.  By
  Lemma~\ref{lem:alternating-quantifiers}, this shows that
  $\onlinewcmk{k}$ is in~$\sigmatwok$, since $\cale$ has a
  polynomial-time solvable winner problem.
The proof of the second statement is analogous.

  The proofs of the third and fourth statements are analogous to the proofs
of the
  third and fourth statements in Theorem~\ref{thm:general-pspace}, except with at
  most $k$ manipulators and reducing from the $\sigmatwok$-complete
  problem $\qbfk{2k}$ rather than from $\qbf$.~\end{proofs}

Note that the (constructive) online manipulation problems considered
in Theorems~\ref{thm:general-pspace} and~\ref{t:pptc} are about
ensuring that the winner set always contains some candidate 
in the $\sigma$ segment stretching from $d$ up to the top-choice.
Now consider ``pinpoint'' variants of
these problems, where we ask whether the distinguished candidate $d$
herself can be guaranteed to be a winner (for nonsequential
manipulation, that version indeed is the one commonly studied).  
Denote the
\emph{pinpoint} variant of, e.g., $\onlinesystemucmk{\cale}{k}$ by
$\pinpointonlinesystemucmk{\cale}{k}$.  Since our hardness proofs in
Theorems~\ref{thm:general-pspace} and~\ref{t:pptc} make all or no one
a winner (and as the upper bounds in these theorems also 
can be seen to hold for the 
pinpoint variants), they establish the corresponding completeness
results also for the pinpoint cases.
We thus have completeness results for $\pspace$ and $\sigmalevel{2k}$
for each $k \geq 1$.  What about the classes $\sigmalevel{2k-1}$
and
$\pilevel{k}$, for each $k \geq 1$?  We can get completeness results
for all these classes by defining appropriate variants of
online manipulation problems.  Let $\mathrm{OMP}$ be any of the
online manipulation problems considered earlier, including the pinpoint variants mentioned
above.  Define $\mathrm{freeform}\hbox{-}\mathrm{OMP}$ to be just as
$\mathrm{OMP}$, except we no longer require the distinguished voter
$u$ to be part of the manipulative coalition---$u$ can be in or can
be out,
and the input must specify, for $u$ and all voters
after~$u$,
which ones are the members of the coalition.  The question of
$\mathrm{freeform}\hbox{-}\mathrm{OMP}$ is whether it is true that for
all actions of the nonmanipulators at or after $u$ 
(for 
specificity as to this problem: if $u$ is a 
nonmanipulator, it will in the input come with 
a preference order) there will be actions (each taken with full information
on cast-before-them votes) of the manipulative coalition members such
that their goal of making some candidate $c$ with $c \geq_{\sigma} d$
(or exactly $d$, in the pinpoint versions) a winner is achieved.
Then, whenever Theorem~\ref{t:pptc} establishes a $\sigmalevel{2k}$ or
$\sigmalevel{2k}$-completeness result for $\mathrm{OMP}$, we obtain a
$\pilevel{2k+1}$ or $\pilevel{2k+1}$-completeness result for
$\mathrm{freeform}\hbox{-}\mathrm{OMP}$
and for $k=0$ manipulators we obtain
$\pilevel{1} = \conp$ or $\conp$-completeness results.
 Similarly, the $\pspace$ and
$\pspace$-completeness results for $\mathrm{OMP}$ we established in
Theorem~\ref{thm:general-pspace} also can be shown true for 
$\mathrm{freeform}\hbox{-}\mathrm{OMP}$.

On the other hand,
if we define a variant of $\mathrm{OMP}$ by
requiring the final voter
to always be a manipulator, the $\pspace$ and
$\pspace$-completeness results for $\mathrm{OMP}$ from
Theorem~\ref{thm:general-pspace} remain true for this variant; the
$\sigmalevel{2k}$ and $\sigmalevel{2k}$-completeness results for
$\mathrm{OMP}$ from Theorem~\ref{t:pptc} change to $\sigmalevel{2k-1}$
and $\sigmalevel{2k-1}$-completeness results for this variant; and the
above $\pilevel{2k+1}$ and $\pilevel{2k+1}$-completeness results for
$\mathrm{freeform}\hbox{-}\mathrm{OMP}$ change to $\pilevel{2k}$ and
$\pilevel{2k}$-completeness results for this variant, $k \geq 1$.

Finally, as an open
direction (and related conjecture),
we define for each of the previously considered variants of
online manipulation problems a \emph{full profile} version.  For
example, for a given election system $\cale$,
$\mathrm{fullprofile}\hbox{-}\onlinesystemucmk{\cale}{k}$ is
the function problem that, given
an OMS \emph{without} any distinguished candidate, $(C, u, V, \sigma)$,
returns a length $\|C\|$ bit-vector that for each candidate $d \in C$
says if the answer to ``$(C, u, V, \sigma, d) \in
\onlinesystemucmk{\cale}{k}$?'' is ``yes'' ($1$) or ``no''
($0$).  The function problem
$\mathrm{fullprofile}\hbox{-}\pinpointonlinesystemucmk{\cale}{k}$ is defined
analogously, except regarding $\pinpointonlinesystemucmk{\cale}{k}$.

It is not hard to prove, as a corollary to Theorem~\ref{t:pptc}, that:
\begin{theorem}
\label{thm:fullprofile-general}
For each election system $\cale$ whose winner problem can be
  solved in polynomial time,
\begin{enumerate}
\item 
  $\mathrm{fullprofile}\hbox{-}\onlinesystemucmk{\cale}{k}$ is in
  $\fp^{\sigmalevel{2k}[\mathcal{O}(\log n)]}$, the class of functions
  computable in polynomial time given Turing access to a
  $\sigmalevel{2k}$ oracle with $\mathcal{O}(\log n)$ queries allowed
  on inputs of size~$n$, and
\item
  $\mathrm{fullprofile}\hbox{-}\pinpointonlinesystemucmk{\cale}{k}$ is
  in $\fp^{\sigmalevel{2k}}_{\mathrm{tt}}$, the class of functions
  computable in polynomial time given truth-table access to a
  $\sigmalevel{2k}$ oracle.
\end{enumerate}
\end{theorem}

We conjecture that both problems are complete for the corresponding
class under metric reductions \cite{kre:j:optimization}, for suitably
defined election systems with polynomial-time winner problems.

If the full profile version of an online manipulation problem can be
computed efficiently, we clearly can also easily solve each of the
decision problems involved by looking at the corresponding bit of the
length $\|C\|$ bit-vector.  Conversely, if there is an efficient
algorithm for an online manipulation decision problem, we can easily
solve its full profile version by running this algorithm for each
candidate in turn.  Thus, we will state our later results only for
online manipulation decision problem.

\begin{proposition}
\label{prop:full-profile}
Let $\mathrm{OMP}$ be any of the online manipulation decision problems
defined above.  Then $\mathrm{fullprofile}\hbox{-}\mathrm{OMP}$ is in
$\fp$ if and only if $\mathrm{OMP}$ is in~$\p$.
\end{proposition}

\section{Results for Specific Natural Voting Systems}

The results of the previous section show that, simply put, even for
election systems with polynomial-time winner problems, online
manipulation can be tremendously difficult.  But what about
\emph{natural} election systems?  We will now take a closer look at
important natural systems.  We will show that online manipulation can
be easy for them, depending on which particular problem is considered,
and we will also see that the constructive and destructive cases can
differ sharply from each other and that it really matters whether we
are in the nonunique-winner model or the unique-winner model.
Finally, in studying the complexity of online manipulation of 
veto elections, we obtain (as 
Theorems~\ref{thm:online-3-candidate-veto-wcm} and~\ref{thm:online-veto-wcm}) what to the best of our knowledge are the 
first 
$\p^{\np[1]}$-completeness and 
$\p^{\np}$-completeness results in the field of 
computational social choice.

\begin{theorem}
\label{thm:fullprofile-cowinner-plurality}
\begin{enumerate}
\item $\onlinesystemwcm{plurality}$ (and thus also
$\onlinesystemucm{plurality}$) is in~$\p$.
\item $\onlinesystemdwcm{plurality}$ (and
thus also $\onlinesystemducm{plurality}$) is in~$\p$.
\end{enumerate}
\end{theorem}

\begin{proofs}
  For the first part, we describe a polynomial-time algorithm for
  $\onlinesystemwcm{plurality}$.  On input $(C, u, V, \sigma, d)$, our
  algorithm checks
  whether one of the candidates in
  $\Gamma_d = \{c \condition c \geq_{\sigma} d\}$ that has the very
  most vote weight so far among the candidates in $\Gamma_d$ would, if
  $u$ and all the manipulators after $u$ voted for her, have at
  least as much vote weight as the total
vote weight 
of the nonmanipulators 
  after $u$ plus the maximum vote weight over all $h \in C$ with $h
  <_{\sigma} d$ currently cast for~$h$.
  This condition can be checked in polynomial time,

  Why is this algorithm correct?  That is, why is it enough to check
  the above condition?
  Suppose this
  condition holds.
  Let $\hat{c}$ be a candidate in $\Gamma_d$
  that has (or ties for) the most current vote weight among the
  candidates in $\Gamma_d$ such that the condition holds
  for~$\hat{c}$.  Then, even if every nonmanipulator after $u$ votes
  for some particular candidate, say~$e$, with $e <_{\sigma} d$, $e$
  cannot have strictly more vote weight than $\hat{c}$, if $u$ and all
  the manipulators after $u$ vote for~$\hat{c}$.  So, with all the
  manipulators from $u$ onward voting for~$\hat{c}$, the only
  candidates who could possibly get strictly more vote weight than
  $\hat{c}$ are ones that are in~$\Gamma_d$, but even if one or more
  of those do, we still have satisfied our goal of making at least one
  candidate that is ``$\geq_{\sigma} d $'' a winner.\footnote{This
    argument does not work in the unique-winner case.}

  Now suppose the above condition is not met.  We must argue that no
  actions of the remaining manipulators---\emph{even ones that may
    depend on intervening behavior of the nonmanipulators}---can
  guarantee that the winner set will always contain some candidate
  in~$\Gamma_d$.  So, the successful action of the remaining
  nonmanipulators after $u$ is the following: Since the above
  condition is not met, there is some candidate~$b$, $b <_{\sigma} d$,
  such that if all nonmanipulators after $u$ vote for $b$ then $b$
  has strictly more vote weight than any candidate~$a \in \Gamma_d$
  would have even if all the manipulators from $u$ onward voted
  for~$a$.  So these nonmanipulators can force the winner set to not
  include any candidate from~$\Gamma_d$, which means that the
  manipulators fail their goal.

  The proof of the second part is similar in flavor.  Suppose we are
  given an input $(C, u, V, \sigma, d)$.
  If $d$ is the
  most preferred candidate in $\sigma$, then our destructive goal is
  impossible to achieve, as someone will always win under plurality,
  thus thwarting the manipulators' goal of having no winners.
  Otherwise (i.e., if $d$ is not the most preferred candidate in~$\sigma$),
\begin{itemize}
\item let $G$ be the maximum current vote weight among all candidates
  in $\{c \condition c >_{\sigma}d\}$ and
\item let $L$ be the maximum current vote weight among all candidates
  in $\{c \condition c \leq_{\sigma}d\}$.
\end{itemize}
  We claim that the manipulators' destructive goal can be
  guaranteed if and only if $G$ plus all the vote weight of $u$ and
  the remaining manipulators after $u$ is strictly greater than $L$
  plus all the vote weight of the remaining nonmanipulators
  after~$u$.  This can easily be evaluated in polynomial time.

  This algorithm is correct, since if this condition holds then we can
  certainly ensure that none of the candidates in
  $\{c \condition c \leq_{\sigma}d\}$
  are winners, as a candidate achieving a current value of $G$ can do
  better than any of them.  And if this condition fails, then the
  nonmanipulators can ignore the manipulators and all vote for a
  candidate in $\{c \condition c \leq_{\sigma}d\}$ currently having a
  vote weight of~$L$, and by doing so will ensure that that candidate
  is a winner.~\end{proofs}

Theorem~\ref{thm:fullprofile-cowinner-plurality}
refers to problems
in the nonunique-winner model.  By contrast, we now show that online
manipulation for weighted plurality voting in the \emph{unique-winner}
model is $\conp$-hard in the \emph{constructive} case and is
$\np$-hard in the \emph{destructive} case.  This is perhaps
the most dramatic, broad contrast yet between the
nonunique-winner model and the unique-winner model, and is the first such contrast
involving plurality.  The key other $\np$-hardness versus $\p$ result
for the nonunique-winner model versus the unique-winner model is due to
Faliszewski, Hemaspaandra, and 
Schnoor~\cite{fal-hem-sch:c:copeland-ties-matter},
but holds only for 
(standard) weighted manipulation for Copeland$^\alpha$ elections ($0 < \alpha < 1$) with
exactly three candidates; for fewer than three both 
cases there are in $\p$ and for more than three
both are $\np$-complete.
In contrast, the $\p$
results of
Theorem~\ref{thm:fullprofile-cowinner-plurality} hold
for all numbers of candidates, and the $\np$-hardness and
$\conp$-hardness results of Theorem~\ref{thm:unique-winner-plurality}
hold whenever there are at least two candidates.

\begin{theorem}
\label{thm:unique-winner-plurality}
\begin{enumerate}
\item The problem $\onlinesystemdwcm{plurality}_{\mathrm{UW}}$ is
  $\np$-hard, even when restricted to only two candidates
(and this also holds when restricted to three, four, ...\ candidates).
\item The problem $\onlinesystemwcm{plurality}_{\mathrm{UW}}$ is
  $\conp$-hard, even when restricted to only two candidates
(and this also holds when restricted to three, four, ...\ candidates).
\end{enumerate}
\end{theorem}

\begin{proofs}
For  the first statement, we prove $\np$-hardness of
$\onlinesystemdwcm{plurality}_{\mathrm{UW}}$
by a reduction from the $\np$-complete
problem $\partition$: Given a nonempty sequence $(w_1, w_2, \ldots , w_z)$ of
positive integers such that $\sum_{i=1}^z w_i = 2W$ for some positive
integer~$W$, does there exist
a set $I \subseteq \{1, 2, \ldots , z\}$ such that $\sum_{i \in
  I} w_i = W$?  Let $m \geq 2$.
 Given an instance $(w_1, w_2,
\ldots , w_z)$ of $\partition$, construct an instance $(\{c_1, \ldots, c_m\},
u_1, V,  c_1 > c_2 > \cdots > c_m, c_1)$ of
$\onlinesystemdwcm{plurality}_{\mathrm{UW}}$ such that
$V$ contains $m+z-2$ voters $v_1, \ldots, v_{m-2}, u_1, \ldots, u_z$
who vote in that order.  For $1 \leq i \leq m-2$, $v_i$ votes for
$c_i$ and has weight $(m-1)W - i$,  and for
$1 \leq i \leq z$, $u_i$ is a manipulator of weight $(m-1)w_i$.
If $(w_1, w_2, \ldots , w_z)$ is a yes-instance of $\partition$,
the manipulators can give  $(m-1)W$ points to both $c_{m-1}$ and~$c_m$,
and zero points to the other candidates.  So $c_{m-1}$ and~$c_m$
are tied for the most points and there is no unique winner.
On the other hand, the only way to avoid having a unique winner
in our $\onlinesystemdwcm{plurality}_{\mathrm{UW}}$  instance is
if there is a tie for the most points.  The only candidates that can
tie are $c_{m-1}$ and $c_m$, since all other pairs of
candidates have different scores modulo $m-1$. 
It is easy to see that $c_{m-1}$ and $c_m$  tie for the most points
only if they both get exactly $(m-1)W$ points.  It follows
that $(w_1, w_2, \ldots , w_z)$ is a yes-instance of
$\partition$.

For the second part, we adapt the above
construction to yield a reduction from $\partition$ to the complement
of $\onlinesystemwcm{plurality}_{\mathrm{UW}}$.
Given an instance $(w_1, w_2, \ldots , w_z)$ of 
$\partition$, construct an instance $(\{c_1, \ldots, c_m\},
\widehat{u}, V,  c_1 > c_2 > \cdots > c_m, c_m)$ of
$\onlinesystemwcm{plurality}_{\mathrm{UW}}$ such that
$V$ contains $m+z-1$ voters $v_1, \ldots, v_{m-2}, \widehat{u},
u_1, \ldots, u_z$ who vote in that order.
For $1 \leq i \leq m-2$, $v_i$ has the same vote and the same weight as
above, $\widehat{u}$ is a manipulator of weight 0, and
for $1 \leq i \leq z$, $u_i$ has the same weight as above, but 
in contrast to the case above, $u_i$ is now a nonmanipulator.
By the same argument as above, it follows that
$(w_1, w_2, \ldots , w_z)$ is a yes-instance of $\partition$ if and
only if the nonmanipulators can ensure that there is no unique winner,
which in turn is true if and only if 
the manipulator can not ensure that there is a unique winner.~\end{proofs}

\begin{theorem}
\label{thm:scoring}
  For each scoring rule $\alpha = (\alpha_1, \ldots , \alpha_m)$,
  $\onlinesystemwcm{\alpha}$ is in $\p$ if $\alpha_2 = \alpha_m$ and is
$\np$-hard otherwise.
\end{theorem}

\begin{proofs}
If $\alpha_1 = \alpha_m$, all candidates are always winners.  If
$\alpha_1 > \alpha_2 = \alpha_m$, this is in essence
$m$-candidate plurality and
$\onlinesystemwcm{plurality}$ is in $\p$ by
Theorem~\ref{thm:fullprofile-cowinner-plurality}.  If $\alpha_2
\neq \alpha_m$, the result follows from the $\np$-hardness of
$\systemwcm{\alpha}$, which is the main theorem from 
Hemaspaandra and Hemaspaandra~\cite{hem-hem:j:dichotomy-scoring},
and Corollary~\ref{cor:offline-to-online}.~\end{proofs}

\begin{theorem}
\label{thm:online-veto-ucm}
For each $k$, $\onlinesystemucm{\mbox{$k$}\hbox{-}approval}$ and
$\onlinesystemucm{\mbox{$k$}\hbox{-}veto}$ are in~$\p$.
\end{theorem}

\begin{proofs}
Consider $1$-veto.
  Given an $\onlinesystemucm{1\hbox{-}veto}$ instance $(C, u, V, \sigma, d)$,
  the best strategy for the manipulators 
{}from $u$ onward (let $n_1$ denote how many of these there are)
is to
  minimize $\max_{c<_{\sigma} d} \score(c)$.  
Let $n_0$ denote how many nonmanipulators come after~$u$.
We claim that $(C, u, V, \sigma, d)$ 
is a yes-instance if and only if $d$ is ranked last in
$\sigma$ or there exists a threshold $t$ 
such that
\begin{enumerate}
\item $\sum_{c<_{\sigma} d} (\maxscore(c) \ominus t) \leq n_1$
  (so those manipulators can ensure that all candidates
  ranked $<_{\sigma} d$ score at most $t$ points),
  where ``$\ominus$'' denotes  proper 
subtraction ($x \ominus y = \max(x-y, 0)$) and
$\maxscore(c)$ is $c$'s score when none of the voters from
$u$ onward veto $c$, and
\item $\sum_{c\geq_{\sigma} d} (\maxscore(c) \ominus (t-1)) >
  n_0$ (so those nonmanipulators cannot prevent that some
  candidate ranked $\geq_{\sigma} d$
  scores at least $t$ points).
\end{enumerate}

For $1$-veto
under the above approach, in each situation where the 
remaining manipulators can force success against all actions of the 
remaining nonmanipulators, $u$ (right then as she moves)
can set her \emph{and all future manipulators'
actions} so as to force
success regardless of the actions of the remaining nonmanipulators.
For $k$-approval and $k$-veto, $k\geq 2$, that approach provably 
cannot work (as will be explained right after this proof); 
rather, we sometimes need later manipulators' actions to be 
shaped by intervening nonmanipulators' actions.
Still, the following P-time algorithm, which works for 
all $k$, tells 
whether
success can be forced.  As a thought experiment, for each voter $v$
from $u$ 
onwards in sequence do this: Order the candidates in 
$\{ c \condition c \geq_\sigma d\}$ from most to least current approvals,
breaking ties arbitrarily,
and postpend the remaining candidates ordered from least to most 
current approvals. 
Let $\ell$ be $k$ for $k$-approval and $\|C\|-k$ for $k$-veto.
Cast the voter's $\ell$ approvals
for the first $\ell$ candidates 
in this order if $v$ is a manipulator, and otherwise for the last $\ell$
candidates in this order.  Success can be forced against perfect
play 
if and only if
this P-time process leads to success.%
{}~\end{proofs}

In the above proof we said that the approach for
$1$-veto (in which the current manipulator can set her and all future
manipulators' actions so as to force success independent of the
actions of intervening future nonmanipulators) provably cannot work
for $k$-approval and $k$-veto, $k\geq 2$.  Why not?  Consider an OMS
$(C, u, V, \sigma, d)$ with candidate set $C = \{c_1, c_2, \ldots ,
c_{2k}\}$, $\sigma$ being given by $c_1 >_{\sigma} c_2 >_{\sigma}
\cdots >_{\sigma} c_{2k}$, and $d=c_1$.  So, $u$'s coalition wants to
enforce that $c_1$ is a winner.  Suppose that $v_1$ has already cast
her vote, now it's $v_2 = u$'s turn, and the order of the future
voters is $v_3, v_4, \ldots , v_{2j}$, where all $v_{2i}$, $2 \leq i
\leq j$, belong to $u$'s coalition, and all $v_{2i-1}$ do not.
Suppose that $v_1$ was approving of the $k$ candidates in $C_1
\subseteq \{c_2, c_3, \ldots , c_{2k}\}$, $\|C_1\| = k$.  Then $u$
must approve of the $k$ candidates in~$\overline{C_1}$, to
ensure that $c_1$ draws level with the candidates in $C_1$ and none of
these candidates can gain another point.  Next, suppose that
nonmanipulator $v_3$ approves of the $k$ candidates in $C_3 \subseteq
\{c_{2}, c_{3}, \ldots , c_{2k}\}$, $\|C_3\| = k$.  Then $v_4$, the
next manipulator, must approve of all candidates in~$\overline{C_3}$,
to ensure that $c_1$ draws level with the candidates in $C_3$ and none
of these candidates can gain another point.  This process is repeated
until the last nonmanipulator, $v_{2j-1}$, approves of the candidates
in $C_{2j-1} \subseteq \{c_{2}, c_{3}, \ldots , c_{2k}\}$,
$\|C_{2j-1}\| = k$, and $v_{2j}$, the final manipulator, is forced to
counter this by approving of all candidates in~$\overline{C_{2j-1}}$,
to ensure that $c_1$ is a winner.  This shows that there can be
arbitrarily long chains such that the action of each manipulator after
$u$ depends on the action of the preceding intervening nonmanipulator.

We now turn to online weighted manipulation for veto when restricted
to three candidates.  We denote this restriction of
$\onlinesystemwcm{veto}$ by $\onlinesystemwcm{veto_{|3}}$.

\begin{theorem}
\label{thm:online-3-candidate-veto-wcm}
  $\onlinesystemwcm{veto_{|3}}$ is~$\p^{\np[1]}$-complete.
\end{theorem}

\begin{proofs}
Let $C=\{a,b,c\}$ and suppose $u$'s
manipulative coalition has the preference order $
a >_{\sigma} b >_{\sigma} c$.  Let $d$ denote the distinguished
candidate.

The $d=c$ inputs of $\onlinesystemwcm{veto_{|3}}$ have a trivial $\p$
algorithm, since all these instances are positive.

Restricted to the $d=a$ inputs, the problem is $\np$-hard, which
follows from $\np$-hardness of $\systemwcm{veto_{|3}}$ due to
Conitzer, Sandholm, and
Lang~\cite{con-lan-san:j:when-hard-to-manipulate} (who note that 
their result is valid in 
the unique-winner and nonunique-winner models)
and
Corollary~\ref{cor:offline-to-online}.  The restricted problem in
this case is also in $\np$, by the
following $\np$ algorithm: Given an instance $(C, u, V, \sigma, a)$ of
$\onlinesystemwcm{veto_{|3}}$ satisfying $a >_{\sigma} b
>_{\sigma} c$, nondeterministically guess a partition
$(A,B)$ of the
manipulators from $u$ onward; all voters in $A$ veto~$b$ and
all voters in $B$ veto~$c$; the nonmanipulators 
after $u$
veto~$a$; on any such path, accept if and only if $a$ is a winner.

Restricted to 
$d=b$, $\onlinesystemwcm{veto_{|3}}$ is $\conp$-hard, which
follows by a reduction from $\partition$ to the complement of
$\onlinesystemwcm{veto_{|3}}$:
Given an instance $(w_1, w_2,
\ldots , w_z)$ of $\partition$, where $\sum_{i=1}^z w_i = 2W > 0$,
construct an instance $(C, u, V, \sigma, d)$ of
$\onlinesystemwcm{veto_{|3}}$ as follows. 
There is one weight
$W-1$ voter before $u$ vetoing~$c$, $u$ has weight zero, and
there are $z$ nonmanipulators after $u$ having weights $w_1, w_2,
\ldots , w_z$.  Now, $(w_1, w_2, \ldots , w_z)$ is a yes-instance of
$\partition$ if and only if $c$ can be made the unique winner by the
nonmanipulators, which in turn is true if and only if there is no
winning strategy for the manipulator to ensure a winner
$\geq_{\sigma} b$.
Observe that in this case
$\onlinesystemwcm{veto_{|3}}$ is also in $\conp$, by
the following $\np$ algorithm for the complement:
Given an instance $(C, u, V, \sigma, b)$ of
$\onlinesystemwcm{veto_{|3}}$ such that $a >_{\sigma} b >_{\sigma}
c$, nondeterministically guess a partition $(A,B)$ of the
nonmanipulators after~$u$; all voters in $A$ veto~$a$ and all
voters in $B$ veto~$b$; the manipulators from $u$ onward veto~$c$;
on any such path, accept if and only if $c$ is the unique winner.

This proves the result, since
$\onlinesystemwcm{veto_{|3}}$ can in light of the above
be written as the union of 
an $\np$-complete and a $\conp$-complete set that are 
P-separable.\footnote{Sets $S_1$ and $S_2$ are said to 
be 
\emph{P-separable}~(see \cite{gro-sel:j:complexity-measures})
if there exists a polynomial-time computable 
set $T$ such that $S_1 \subseteq T \subseteq \overline{S_2}$.
(One cannot in our main text 
change ``are P-separable'' into ``are disjoint,'' as
then the 
reasoning used would become invalid;
for example, SAT and $\rm \overline{\mathrm{SAT}}$
are disjoint, and are respectively NP- and coNP-complete, but their 
union is $\Sigma^\star$ and so unless $\p = \np$ will not be 
$\p^{\np[1]}$-complete.)}%
{}~\end{proofs}

Moving from three to four candidates increases 
the complexity, namely to $\p^\np$-completeness, and 
that same bound holds for unlimitedly many candidates.
Although this is a strict increase in complexity from 
$\p^{\np[1]}$-completeness
(unless the polynomial hierarchy
collapses~\cite{kad:joutdatedbychangkadin:bh,for-pav-sen:j:proving-sat-does-not-have-small-circuits}),
membership in $\p^\np$ still places 
this problem far below the general PSPACE bound
from earlier in this paper.

\begin{theorem}
\label{thm:online-veto-wcm}
  $\onlinesystemwcm{veto}$ is~$\p^{\np}$-complete, even when restricted
to only four candidates.
\end{theorem}

\begin{proofs}
We first show that $\onlinesystemwcm{veto}$ is in~$\p^{\np}$.
The proof is reminiscent of the proof for 1-veto in
Theorem~\ref{thm:online-veto-ucm}.  Let $(C, u, V, \sigma, d)$ be a
given instance of $\onlinesystemwcm{veto}$ with $C = \{c_1, c_2,
\ldots , c_m\}$ and $c_1 >_{\sigma} c_2 >_{\sigma} \cdots >_{\sigma}
c_m$.  Suppose $d = c_i$.  Our $\p^{\np}$ algorithm proceeds as
follows:

\begin{enumerate}
\item Compute the minimal threshold $t_1$ such that there exists a
  partition $(A_{i+1}, \ldots , A_m)$ of the weights of the
  manipulators from $u$ onward such that for each~$j$, $i+1 \leq j
  \leq m$,
\[
\maxscore(c_j) - \sum A_j \leq t_1,
\]
where $\maxscore(c_j)$ is $c_j$'s score when none of the voters from
$u$ onward veto $c$.
That is, by having manipulators from $u$ onward with weights in
$A_j$ veto $c_j$, the manipulators from $u$ onward can ensure that
none of the candidates they dislike more than $d$ exceeds a score
of~$t_1$.

\item Compute the minimal threshold $t_2$ such that there exists a
  partition $(A_1, \ldots , A_i)$ of the weights of the
  nonmanipulators after $u$ such that for each~$j$, $1 \leq j \leq i$,
\[
\maxscore(c_j) - \sum A_j \leq t_2.
\]
That is, if the nonmanipulators after $u$ with weights in $A_j$
veto $c_j$, none of the candidates that the manipulators like as least
as much as $d$ exceeds a score of~$t_2$.

\item Accept if and only if $t_1 \leq t_2$.
\end{enumerate}
Note that the first two steps of the algorithm can both be done in
$\fp^{\np}$ by using an $\np$ oracle that checks whether
there exists a partition of the specified kind.

It remains to show that $\onlinesystemwcm{veto_{|4}}$ is~$\p^{\np}$-hard.
We will reduce from the standard $\p^{\np}$-complete problem 
MAXSATASG$_=$, which is the set of pairs of 3cnf formulas\footnote{We
  denote a formula in conjunctive normal form by \emph{cnf formula},
  and a \emph{3cnf formula} is a cnf formula with exactly three literals
  per clause.}
that have the same maximal satisfying 
assignment~\cite{wag:j:more-on-bh}.
\ehnote{Wagner only states the completeness, and he doesn't define what
happens if the formulas are not satisfiable. 
Also, his variables start at $x_0$ and $x_0$ is the least significant bit.
And he simply gives his
formulas as $\phi(x_0,x_1, \ldots, x_n)$ 
which technically is not correct.}%
To be precise, we will assume that our propositional variables
are $x_1, x_2, \ldots$.  If $x_n$ is the largest propositional
variable occurring in $\phi$, we often
write $\phi(x_1, \ldots, x_n)$ to make that explicit.
An assignment for $\phi(x_1, \ldots, x_n)$ is an 
$n$-bit string $\alpha$ such that $\alpha_i$ gives the assignment
for variable $x_i$.   We will
sometimes identify $\alpha$ with the binary integer
it represents.
For $\phi$ a formula, $\msa(\phi)$ is the lexicographically largest
satisfying assignment
for $\phi$.  If $\phi$ is not satisfiable, $\msa(\phi)$ is not defined.
And we define MAXSATASG$_=$ as the set of pairs of 3cnf formulas
$(\phi(x_1, \ldots, x_n),\psi(x_1, \ldots, x_n))$ such that
$\phi$  and $\psi$ are satisfiable 3cnf formulas, and
$\msa(\phi) = \msa(\psi)$.

The OMS that we will construct will have four candidates, 
$a >_\sigma b >_\sigma c >_\sigma d$, and the distinguished candidate
will be $b$.  Looking at the~$\p^{\np}$ algorithm above, we can
see that determining whether the OMS can be manipulated basically amounts to
determining whether the nonmanipulator weights have a ``better''
partition than the manipulator weights. 

So, we will associate formulas with multisets of positive
integers, and their satisfying assignments with subset sums.
This already happens in the standard reduction from 3SAT to
SUBSETSUM.  However, we also want larger satisfying assignments
to correspond to ``better'' subset sums.  In order to do this,
we use Wagner's variation of the 3SAT to SUBSETSUM
reduction~\cite{wag:j:more-on-bh}.  Wagner uses this reduction to
prove that determining whether the largest subset sum up to a certain
bound is odd is a $\p^\np$-hard problem.

\begin{lemma}
\label{lem:wagner}
Let $\phi(x_1, \ldots,x_n)$ be a 3cnf formula.
Wagner's reduction
maps this formula to an instance $(k_1, \ldots, k_t, L)$ of
SUBSETSUM with the following properties:
\begin{enumerate}
\item For all assignments $\alpha$, $\phi[\alpha]$ if and only if
there exists a subset of $k_1, \ldots, k_t$ that sums to $L+\alpha$.
\item For all $K$ such that $2^n \leq K \leq 2(2^n-1)$,
no subset of $k_1, \ldots, k_t$ sums to $L + K$.
\end{enumerate}
\end{lemma}

\sproofof{Lemma~\ref{lem:wagner}}
The first claim is immediate from the proof of Theorem 8.1(3)
from~\cite{wag:j:more-on-bh}.
\ehnote{Wagner uses $n+1$ variables instead of $n$ and there are a couple
of minor typos in the construction.  In particular, ``where
$b_{ij} = 1$ if $x_i$ is in clause $j$'' should be 
``where $b_{ij} = 1$ if $\overline{x_i}$ is in clause $j$'' and
in the definition of $L$, ``$m$th digit'' should be ``$(m+1)$st digit''
and ``$(m+n+1)$th digit'' should be ``$(m+n+2)$nd digit.''

Wagner's max sat asg is more or less the reverse of mine.
}%
For the second claim, note that
$L + K \leq L + 2(2^n-1) < L + 6^n$.
In Wagner's construction, 
$L = \underbrace{3 \cdots 3}_m \underbrace{1 \cdots 1}_n
\underbrace{0 \cdots 0}_n$ in base 6, where $m$ is the number of clauses
in $\phi$.
So, $(L + K)$'s representation base 6 is
$\underbrace{3 \cdots 3}_m \underbrace{1 \cdots 1}_n$ followed by
$n$ digits.
It is easy to see from Wagner's construction
that the subset sums of this form that can be realized
are exactly $L + \beta$, where $\beta$ is a satisfying
assignment of $\phi$.  Since $K \geq 2^n$,
$K$ is not even an assignment, and thus
no subset of $k_1, \ldots, k_t$ sums to
$L + K$.~\eproofof{Lemma~\ref{lem:wagner}}

Let $\phi(x_1, \ldots, x_n)$ and $\psi(x_1, \ldots, x_n)$ be
3cnf formulas, and consider instance $(\phi,\psi)$
of MAXSATASG$_{=}$.
Without loss of generality, we assume that $x_1$ does not
actually occur in
$\phi$ or $\psi$.
We will define an OMS
$(C, u, V, \sigma, b)$ with $C = \{a,b,c,d\}$
and $\sigma = a > b > c > d$ such that $(\phi,\psi) \in $
MAXSATASG$_{=}$ if and only if $(C, u, V, \sigma, b)$
is a positive instance of $\onlinesystemwcm{veto}$.
Note that MAXSATASG$_{=}$ corresponds to optimal solutions being equal,
while $\onlinesystemwcm{veto}$ corresponds to one optimal solution being
at least as good as the other.   We will first modify the formulas
such that we also look at the optimal solution for one formula being
at least as good as the optimal solution for the other.
The following is immediate.

\begin{claim}
\label{msa-one}
$(\phi,\psi) \in $ MAXSATASG$_{=}$ if and only if 
$\phi \wedge \psi$ is satisfiable and
$\msa(\phi \wedge \psi) \geq  \msa(\phi \vee \psi)$.
\end{claim}

It will also be very useful if one of the formulas is always 
satisfiable.  We can easily ensure this by adding an extra variable
that will correspond to the highest order bit of the satisfying assignment.
Recall that $x_1$ does not occur in $\phi$ or $\psi$.

\begin{claim}
\label{msa-two}
$(\phi,\psi) \in $ MAXSATASG$_{=}$ if and only if 
$\phi \wedge \psi \wedge x_1$ is satisfiable and
\[
\msa(\phi \wedge \psi \wedge x_1) \geq
\msa(\phi \vee \psi \vee \overline{x_1}).
\]
\ehnote{Displaymath is just for TARK lay-out. Change back
to in-line in TR.}%
\end{claim}

Now we would like to apply the reduction from Lemma~\ref{lem:wagner} on
$\phi \wedge \psi \wedge x_1$ and $\phi \vee \psi \vee \overline{x_1}$.
But wait!
This reduction is defined for 3cnf formulas, and 
$\phi \vee \psi \vee \overline{x_1}$ is not in 3cnf.  Since
$\phi$ and $\psi$ are in 3cnf, it is easy to convert 
$\phi \vee \psi \vee \overline{x_1}$ into cnf in polynomial time.
Let $g$ be the standard reduction from CNF-SAT to 3SAT.  
We can rename the variables such that $g$ has the following property:
For $\xi(x_1, \ldots,x_n)$ a cnf formula,
$g(\xi)(x_1, \ldots, x_n, x_{n+1}, \ldots, x_{\hat{n}})$ is a 3cnf
formula such that $\hat{n} > n$ and such that
for all assignments $\alpha \in \{0,1\}^n$,
$\xi[\alpha]$ if and only if there exists an assignment $\beta \in
\{0,1\}^{\hat{n}-n}$ such that $g(\xi)[\alpha\beta]$.

Let $\widehat{\psi}(x_1, \ldots, x_{\hat{n}}) =
g(\phi \vee \psi \vee \overline{x_1})$.
Let $\widehat{\phi}(x_1, \ldots, x_{\hat{n}}) =
\phi \wedge \psi \wedge (x_1 \vee x_1 \vee x_1)
\wedge (x_{\hat{n}} \vee x_{\hat{n}} \vee \overline{x_{\hat{n}}})$.

\begin{claim}
\label{cl:msa-three}
\begin{itemize}
\item $\widehat{\phi}$ and $\widehat{\psi}$ are in 3cnf and
$\widehat{\psi}$ is satisfiable.
\item
$(\phi,\psi) \in $ MAXSATASG$_{=}$ if and only if 
$\widehat{\phi}$ is satisfiable and
$\msa(\widehat{\phi}) \geq \msa(\widehat{\psi})$.
\end{itemize}
\end{claim}

\sproofof{Claim\ref{cl:msa-three}}
From the previous claim we know that
if $(\phi,\psi) \in $ MAXSATASG$_{=}$, then
$\phi \wedge \psi \wedge x_1$ is satisfiable and
thus $\widehat{\phi}$ is satisfiable.
Also from the previous claim, if $(\phi,\psi) \in $ MAXSATASG$_{=}$, then
$\msa(\phi \wedge \psi \wedge x_1) \geq
\msa(\phi \vee \psi \vee \overline{x_1})$.
Let $\alpha$ be the maximal satisfying assignment of
$\phi \wedge \psi \wedge x_1$.  Then $\alpha 1^{\hat{n} - n}$ is
the maximal satisfying assignment of $\widehat{\phi}$.
Let $\alpha'$ be the maximal satisfying assignment of
$\phi \vee \psi \vee \overline{x_1}$.
Then $\alpha' \beta$ is the maximal satisfying assignment of $\widehat{\psi}$
for some $\beta$.  Since $\alpha \geq \alpha'$, it follows that
$\alpha 1^{\hat{n} - n} \geq \alpha' \beta$.  

For the converse, suppose that
$\widehat{\phi}$ is satisfiable and
$\msa(\widehat{\phi}) \geq
\msa(\widehat{\psi})$.  Let $\gamma$ be the 
maximal satisfying assignment of $\widehat{\phi}$ and
let $\gamma'$ be the maximal satisfying assignment 
of $\widehat{\psi}$.  Then the length-$n$ prefix of $\gamma$
is the maximal satisfying assignment of $\phi \wedge \psi \wedge x_1$
and the length-$n$ prefix of $\gamma'$
is the maximal satisfying assignment of
$\phi \vee \psi \vee \overline{x_1}$.
Since $\gamma \geq \gamma'$, the $n$-bit prefix of $\gamma$ is
greater than or equal to the $n$-bit prefix of
$\gamma'$.~\eproofof{Claim\ref{cl:msa-three}}

We now apply Wagner's reduction
from Lemma~\ref{lem:wagner} to $\widehat{\phi}$ and $\widehat{\psi}$.
Let $k_1, \ldots, k_t, L$ be the output of Wagner's reduction
on $\widehat{\phi}$ and let
$k'_1, \ldots, k'_{t'}, L'$ be the output of Wagner's reduction on
$\widehat{\psi}$.

As mentioned previously, we will define an OMS
$(C, u, V, \sigma, b)$ with $C = \{a,b,c,d\}$
and $\sigma = a > b > c > d$ such that $(\phi,\psi) \in $
MAXSATASG$_{=}$ if and only if $(C, u, V, \sigma, b)$ 
is a positive instance of $\onlinesystemwcm{veto}$.
Because we are looking at veto, when determining the outcome
of an election, it is easiest to simply count the number of
vetoes for each candidate.  Winners have the fewest vetoes.  
For $\hat{c}$ a candidate, we will denote the total weight of the voters
that veto $\hat{c}$ by $\vetoes(\hat{c})$.

There are four voters in $V_{<u}$: one voter of weight
$L$ vetoing $a$, one voter of
weight $L + 2L' + 2(2^{\hat{n}}-1) - \sum k'_i$ vetoing $b$,
one voter of weight $L'$ vetoing $c$, and one voter of weight
$L' + 2L + 2(2^{\hat{n}}-1) - \sum k_i$ vetoing $d$.
Let $u = u_1$.
$V_{u<}$ consists of $t-1$ further manipulators $u_2, \ldots, u_t$ followed
by $t'$ nonmanipulators $u'_1, \ldots, u'_{t'}$.
The weight of manipulator $u_i$ is $k_i$ and the weight of nonmanipulator
$u'_i$ is $k'_i$.

It remains to show that the reduction is correct.  First suppose
that $(\phi,\psi)$ is in MAXSATASG$_{=}$.  By 
Claim~\ref{cl:msa-three},  this implies that
$\widehat{\phi}$ and $\widehat{\psi}$ are satisfiable 3cnf formulas such that
$\msa(\widehat{\phi}) \geq \msa(\widehat{\psi})$.
Let $\alpha =  \msa(\widehat{\phi}$).
We know from Lemma~\ref{lem:wagner} that there exists a subset
of $k_1, \ldots, k_t$ that sums to $L + \alpha$.
The manipulators corresponding to this subset will veto $c$,
so that $c$ receives $L + \alpha$ vetoes from the manipulators. 
The remaining manipulators will veto $d$, i.e., $d$ receives
$\left(\sum k_i\right) - L - \alpha$ vetoes from the manipulators.
After the manipulators
have voted, $\vetoes(a) = L, \vetoes(b) = L + 2L' + 2(2^{\hat{n}}-1) - \sum k'_i$,
$\vetoes(c) = L' + L + \alpha$, and $\vetoes(d) = L' + L + 2(2^{\hat{n}}-1) - \alpha$.
Since $\alpha \leq 2^{\hat{n}} - 1$, $\vetoes(c) \leq \vetoes(d)$.
We will show that no matter how the nonmanipulators vote,
$a$ or $b$ is a winner.  Suppose for a contradiction that after the
nonmanipulators have voted, $\vetoes(a) > \vetoes(c)$ and
$\vetoes(b) > vetoes(c)$.  If that were
to happen, there would be a subset of $k'_1, \ldots, k'_{t'}$ 
summing to $K$ such that
$L + K = \vetoes(a) >  \vetoes(c) = L + L' + \alpha$ and
$L + 2L' + 2(2^{\hat{n}}-1) - K = \vetoes(b) > \vetoes(c) = L + L' + \alpha$.
It follows that $\alpha < K - L' < 2(2^{\hat{n}}-1)$ and
there exists a subset of $k'_1, \ldots, k'_{t'}$ that sums to
$L' + (K - L')$.  It follows from Lemma~\ref{lem:wagner} that
$K-L'$ is a satisfying assignment for $\widehat{\psi}$.  But that
contradicts the assumption that $\msa(\widehat{\phi}) \geq
\msa(\widehat{\psi})$.

The proof of the converse is very similar.
\ehnote{In fact, it is so similar that a well-chosen claim that
states the relationship between assignments for $\widehat{\phi}$ and
$\max \min (\vetoes(c), \vetoes(d))$ and
between assignments for $\widehat{\psi}$ and
$\max \min (\vetoes(a), \vetoes(b))$ could
probably reduce the length of the proof (though there
is a slight asymmetry between $\widehat{\phi}$ and $\widehat{\psi}$,
since the latter is always satisfiable).
But let's leave that for after the TARK submission.}%
Suppose that $(\phi,\psi) \not \in $ MAXSATASG$_{=}$.
By Claim~\ref{cl:msa-three},  $\widehat{\psi}$ is satisfiable.
Let $\alpha = \msa(\widehat{\psi})$.  By 
Claim~\ref{cl:msa-three},  either
$\widehat{\phi}$ is not satisfiable or
$\msa(\widehat{\phi}) < \alpha$.
Suppose the manipulators vote such that $c$ receives $K$ vetoes from
some of them.  Without loss of generality, assume all other
manipulators veto $d$, so that $d$ receives $\left(\sum k_i\right) - K$
vetoes from the manipulators.
We know from Lemma~\ref{lem:wagner} that there exists a subset
of $k'_1, \ldots, k'_{t'}$ that sums to $L' + \alpha$.
After the manipulators have voted, the nonmanipulators will vote
such that $a$ receives $L' + \alpha$ vetoes from the nonmanipulators
and the remaining nonmanipulators will veto $b$, i.e., $b$ receives
$\left(\sum k'_i\right) - L' - \alpha$ vetoes from the nonmanipulators. 
So, $\vetoes(a) = L + L' + \alpha$,
$\vetoes(b) = L + L' + 2(2^{\hat{n}}-1) - \alpha$,
$\vetoes(c) = L' + K$,
and $\vetoes(d) = L' + 2L + 2(2^{\hat{n}}-1) - K$.
We will show that neither $a$ nor $b$ is a winner.
Since $\alpha \leq 2^{\hat{n}} - 1$, $\vetoes(a) \leq \vetoes(b)$.
So it suffices to show that $a$ is not a winner.  If 
$a$ were a winner, $\vetoes(a) \leq \vetoes(c)$ and 
$\vetoes(a) \leq \vetoes(d)$.
This implies that
$\alpha \leq K - L \leq 2(2^{\hat{n}} - 1)$.
It follows from Lemma~\ref{lem:wagner} that
$K-L$ is a satisfying assignment for $\widehat{\phi}$.  But that
contradicts the assumption that either
$\widehat{\phi}$ is not satisfiable or
$\msa(\widehat{\phi}) < \alpha$.~\end{proofs}

{}Immediately from
Theorems~\ref{thm:online-veto-ucm}
and~\ref{thm:online-veto-wcm},
we have that 
the full profile variants of $\onlinesystemucm{\mbox{$k$}\hbox{-}veto}$ and
$\onlinesystemucm{\mbox{$k$}\hbox{-}approval}$ are in~$\fp$ and that
$\mathrm{fullprofile}\hbox{-}\onlinesystemwcm{veto}$ is
in~$\fp^{\np}$.

\section{Uncertainty About the Order of Future Voters}

So far, we have been dealing with cases where 
the order of future voters was fixed and known.  
But what happens if the order of future voters itself 
is unknown?  Even here, we can make claims.  To model this 
most naturally, our ``magnifying-glass moment'' will
focus not on one manipulator $u$, but will focus at 
a moment in time when some voters 
are still to come (as before, we know who they are and 
which are manipulators;  as before, we have a preference
order $\sigma$, and know what votes have been cast so 
far, and have a distinguished candidate $d$).
And the question our problem is 
asking is:  Is it the case that 
our manipulative coalition can ensure 
that the winner set will 
always
include $d$ or someone 
liked more than $d$ with respect to $\sigma$ 
(i.e., the winner set will have nonempty intersection 
with 
$\{c \in C
\condition c \geq_{\sigma} d\}$),
\emph{regardless of what order the remaining 
voters vote in}. 
We will call this problem the 
\emph{schedule-robust online
  manipulation problem}, and will
denote it by $\mathrm{SR}\hbox{-}\onlineucm$.
(We will add a ``[1,1]'' suffix for the 
restriction of this problem to instances when
at most one manipulator and at most one 
nonmanipulator have not yet
voted.)
One might 
think that this problem captures both a 
$\sigmatwo$ and a $\pitwo$ issue, and 
so would be hard for both classes.  
However, the requirement of schedule robustness
tames the problem
(basically what underpins that is simply that 
exists-forall-predicate implies 
forall-exists-predicate), 
bringing it into
$\sigmatwo$.  Further, we can prove, by 
explicit construction of such a system, that for 
some simple election 
systems this problem is 
complete for $\sigmatwo$.

\begin{theorem}
\label{thm:schedulerobust-sigmatwo}
\begin{enumerate}
\item For each election system $\cale$ whose winner problem is in~$\p$,
  $\schedulerobustonlinesystemucm{\cale}$ is in~$\sigmalevel{2}$.

\item There exists an election system $\cale$, whose winner problem is
  in~$\p$, such that 
$\schedulerobustonlinesystemucm{\cale}$ 
(indeed, even $\schedulerobustonlinesystemucmk{\cale}{1,1}$) 
is $\sigmalevel{2}$-complete.
\end{enumerate}
\end{theorem}

\begin{proofs}
  For the first part, note that for each $\p$ predicate~$R$, each
  polynomial~$p$, and each string~$x$, we have that
\begin{equation}
\label{eq:exists-forall-implies-forall-exists}
(\exists^p y)\, (\forall^p z)\, [R(x,y,z)] \Rrightarrow
(\forall^p z)\, (\exists^p y)\, [R(x,y,z)].
\end{equation}
Given an input of $\schedulerobustonlinesystemucm{\cale}$, we have to
decide whether, regardless of the order of the future voters, the
manipulative coalition can ensure that the winner set will always
include the distinguished candidate $d$ or someone liked more than $d$
with respect to its preference order.  Note that manipulators
correspond to existential quantifiers and nonmanipulators correspond
to universal quantifiers.
By~(\ref{eq:exists-forall-implies-forall-exists}), among any two fixed
orders of future voters where in the first order some manipulator
precedes some nonmanipulator and in the second they are swapped
(everything else being the same), the former is the more demanding
one.  Thus, by repeatedly
applying~(\ref{eq:exists-forall-implies-forall-exists}), any order of
future voters that has all remaining manipulators first, followed by
all remaining nonmanipulators, will be most demanding for the
manipulators.  Since schedule robustness requires the manipulators to
force success for \emph{all possible} orders of future voters, it is
enough to require them to force success for such a ``most demanding
order.''  Since $\cale$ winners can be determined in polynomial time,
this shows that testing whether the manipulative coalition is
successful can be expressed as a $\sigmalevel{2}$ predicate.

For the second part, we define an election system $\cale$ as follows.
Let $R$ be  a $\p$ predicate such that the set
\[
L_2 = \{ x \condition (\exists y : |y|=|x|)\, (\forall z : |z|=|x|)\,
[R(x,y,z)]
\]
is $\sigmalevel{2}$-complete.  Let $(C,V)$ be a given election.
Similar to the definition of the election system in the proof of
Theorem~\ref{thm:general-pspace}, the lexicographically least
candidate in $C$ will specify $x$ by her name; the lexicographically
least voter in $V$ will specify $y$ by her vote; and the
lexicographically greatest voter in $V$ will specify $z$ by her vote.
If there are not enough candidates in $C$ to have $y$ and $z$ of
length $|x|$ (in our fixed encoding of votes), everyone loses
in~$\cale$.  Otherwise, if $R(x,y,z)$ holds then everyone wins
in~$\cale$, else everyone loses in~$\cale$.  This completes the
specification of election system~$\cale$.
Since $R$ is in~$\p$, $\cale$ has a polynomial-time winner
problem.  

The upper bound, $\schedulerobustonlinesystemucm{\cale} \in
\sigmalevel{2}$, follows immediately from the first part.  For the
lower bound, we now define a $\manyonereducesto$-reduction from the
$\sigmalevel{2}$-complete problem $L_2$ to
$\schedulerobustonlinesystemucm{\cale}$ (indeed, even to
$\schedulerobustonlinesystemucmk{\cale}{1,1}$), showing
$\sigmalevel{2}$-hardness of the problem.  Given an instance $x$ of
$L_2$ to an instance $(C, u, V, \sigma, d)$ of
$\schedulerobustonlinesystemucmk{\cale}{1,1}$ as follows: $C$ contains
$x$ as its lexicographically least candidate and enough dummy
candidates, each with a greater name than~$x$; $V$ contains two
voters, a manipulator $u$ and a nonmanipulator $v$ (with $u$'s name
lexicographically smaller than $v$'s); the preference order $\sigma$
is irrelevant, so we fix any order; and it is also irrelevant which
candidate is the distinguished candidate, since all or no one wins
in~$\cale$, so we fix any candidate~$d$.

If $x \in L_2$, and $u$ casts a $y$, $|y|=|x|$, witnessing that (i.e.,
$u$ casts a $y$ such that for each~$z$, $|z|=|x|$, $R(x,y,z)$), then
$d$ wins if $u$ casts $y$ before $v$ casts $z$, and
the same $y$ makes
$d$ win if $v$ casts $z$ before $u$ casts $y$.  If $x \not\in L_2$,
then no matter which of $u$ and $v$ casts her vote first, no $y$ cast
by $u$ can make $d$ win.~\end{proofs}

\section{Conclusions and Open Questions}

We introduced a novel framework for online manipulation in 
sequential voting, and showed that manipulation there can be 
tremendously complex even for systems with simple winner 
problems.   We also showed that among the most important 
election systems, some have 
efficient
online manipulation algorithms but others 
(unless $\p = \np$) do not.
It will be important to, complementing our work,
conduct typical-case
complexity studies.  
We have extended the scope of our
investigation by studying online 
control~\cite{hem-hem-rot:c:online-voter-control,hem-hem-rot:c:online-candidate-control}
and will 
also study online bribery.

\paragraph{Acknowledgments}
We thank COMSOC-2012 reviewers for helpful comments.

\newcommand{\etalchar}[1]{$^{#1}$}

 \bibliographystyle{alpha}

\end{document}